\documentclass[12pt]{article}
\usepackage{graphicx,amsmath}
\usepackage{units}

\newlength{\dinwidth}
\newlength{\dinmargin}
\def\lapproxeq{\lower .7ex\hbox{$\;\stackrel{\textstyle                                                    
<}{\sim}\;$}}                                                    
\def\gapproxeq{\lower .7ex\hbox{$\;\stackrel{\textstyle                                                    
>}{\sim}\;$}}                                                    
\def\be{\begin{equation}}                                                    
\def\ee{\end{equation}}                                                    
\def\bea{\begin{eqnarray}}                                                    
\def\eea{\end{eqnarray}}

\def\qq{q\bar{q}}

\def\GeV{\rm GeV}

\def\sh{\hat s}
\def\sh2{{\hat s}^2}
\def\PDF{{\rm PDF}}
\def\CLO{C^{\rm LO}}
\def\CNLO{C^{\rm NLO}}
\def\CNNLO{C^{\rm NNLO}}
\begin{document}
%\titlepage

\begin{flushright}                                                    
IPPP/12/87  \\
DCPT/12/174 \\                                                    
\today \\                                                    
\end{flushright} 

\vspace*{0.5cm}

\begin{center}
{\Large \bf The LHC can probe small $x$ PDFs;}\\
\vspace*{0.3cm}

{\Large \bf the treatment of the infrared region}

\vspace*{1cm}
                                                   
A.D. Martin$^a$, E.G. de Oliveira$^{a,b}$ and M.G. Ryskin$^{a,c}$  \\                                                    
                                                   
\vspace*{0.5cm}                                                    
$^a$ Institute for Particle Physics Phenomenology, University of Durham, Durham, DH1 3LE \\                                                   
$^b$ Instituto de F\'{\i}sica, Universidade de S\~{a}o Paulo, C.P.
66318,05315-970 S\~{a}o Paulo, Brazil \\
$^c$ Petersburg Nuclear Physics Institute, NRC Kurchatov Institute, Gatchina, St.~Petersburg, 188300, Russia \\          
                                                    
\vspace*{1cm}                                                    
                                                    
\begin{abstract}                                                    

First, we show how to reduce the sensitivity of the NLO predictions of the Drell-Yan production of low-mass, lepton-pairs, at high rapidity, to the choice of factorization scale.  In this way, observations of this process at the LHC can make direct measurements of parton distribution functions in the low $x$ domain; $x \lapproxeq 10^{-4}$. Second, we find an inconsistency in the conventional NLO treatment of the infrared region. We illustrate the problem using the NLO coefficient function of Drell-Yan production. 

\end{abstract}                                                        
\vspace*{0.5cm}                                                    
                                                    
\end{center}

\section{LHC Drell-Yan production as a probe of low $x$}

The very high energy of the LHC allows a probe of the parton distribution functions (PDFs) of the proton at extremely small $x$, a region not accessible at previous accelerators.  To extract the PDFs we describe the experimentally observed cross sections as a convolution of the PDFs and the cross section for the hard partonic subprocess, which is of the form
\be
\label{eq:sig}
d\sigma/d^3p~=~\int dx_1dx_2~\PDF(x_1,\mu_F)~|{\cal M}(p;\mu_F,\mu_R)|^2~\PDF(x_2,\mu_F)\ ,
\ee
where a sum over the various pairs of PDFs is implied. 

\begin{figure} [htb]
\begin{center}
\includegraphics[height=7cm]{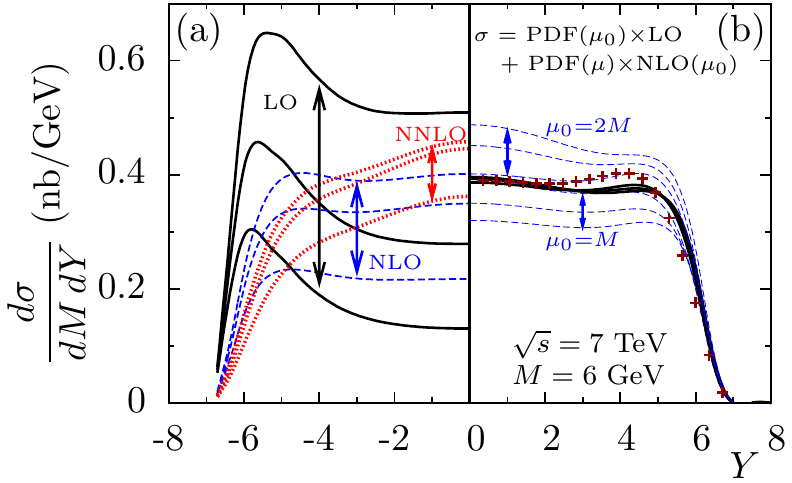}
\vspace{-0.2cm}
\caption{\sf (a) Sensitivity of $M=6$ GeV Drell-Yan $\mu^+\mu^-$ production, as a function of rapidity $Y$, to the choice of factorization scale: $\mu_F=M/2,\ M,\ 2M$, at LO, NLO, NNLO. (b) The bold lines correspond to the choice $\mu_F=\mu_0=1.4M$ which minimizes $\CNLO_{\rm rem}$, and show the stability with respect to the variations $\mu=M/2,\ M,\ 2M$ in the scale of the PDFs convoluted with $\CNLO_{\rm rem}$ -- the $\mu$ dependence is indicated by the symbolic equation at the top of the diagram. The dashed lines show that the stability disappears for other choices of $\mu_0$. The small crosses are the NNLO result. In this figure, taken from \cite{OMR}, the renormalization scale is fixed at $\mu_R=M$.}
\label{fig:muF}
\end{center}
\end{figure}

Here we focus on Drell-Yan production of a low mass $\mu^+\mu^-$ pair. At LO the production of a $\mu^+\mu^-$ system of mass $M$ and rapidity $Y$ arises from the subprocess $\gamma^* \to \qq$ with 
$x_{1,2}=M\exp(\pm Y)/\sqrt{s}$. So for $M=6$ GeV, $Y=4$  (a domain which is accessible to the LHCb experiment) we probe
$x_1=4.7\times 10^{-2},~x_2=1.6\times 10^{-5}$.  The problem is that, in the low $x$ region, the PDFs strongly depend on the choice of the factorization scale $\mu_F$, see Fig. \ref{fig:muF}(a). It is made worse due to the dominance of the
gluon PDF at small $x$, which means that the LO $q\bar{q} \to \gamma^*$ subprocess is overshadowed by
the NLO subprocess $gq \to q\gamma^*$.  However, it is this very dominance
which will allow us to introduce a procedure which greatly suppresses
the scale dependence of the predictions.

The plan is to choose the value of $\mu_F$ which minimizes the higher order $\alpha_s$ NLO, NNLO,.. contributions.
To sketch the idea, we start with the LO expression for the cross section:
$\sigma(\mu_F)=\PDF(\mu_F)\otimes\CLO \otimes \PDF(\mu_F)$ in the collinear approach.
The effect of varying the scale from $m$ to $\mu_F$, in both the left and right PDFs, can be expressed, to first order in $\alpha_s$, as
\be 
\sigma(\mu_F)=\PDF(m)\otimes    
 \left(\CLO ~+~\frac{\alpha_s}{2\pi}{\rm ln}\left(\frac{\mu_F^2}{m^2}\right)(P_{\rm left}\CLO+\CLO P_{\rm right})\right)\otimes \PDF(m),
\label{eq:5}
\ee
where the splitting functions
$P_{\rm right}=P_{qq}+P_{qg}$ and $P_{\rm left}=P_{\bar{q}\bar{q}}+P_{\bar{q}g}$
act on the right and left PDFs respectively. We may equally well have incoming $\bar{q}$'s in $P_{\rm right}$ and incoming $q$'s in $P_{\rm left}$.

  At NLO we may write
\be
\sigma(\mu_F)~=~\PDF(\mu_F)\otimes(\CLO + \alpha_s \CNLO_{\rm corr}) \otimes \PDF(\mu_F), 
\label{eq:stab}
\ee
where the $2\to 2$ subprocesses $\qq \to g\gamma^*$ and  $gq \to q\gamma^* $ are now calculated with better, than LLA, accuracy. However, we must subtract from $\CNLO$, the part of the contribution already included, to LLA accuracy, in the $\alpha_s$ term in (\ref{eq:5}).  The remaining contribution, $\CNLO_{\rm rem}(\mu_F)$, now depends on the scale $\mu_F$, coming from the $\mu_F$ dependence of the LO LLA term that has been subtracted off. The trick is to choose an appropriate scale, $\mu_F=\mu_0$, so as to minimize the remaining NLO contribution $\CNLO_{\rm rem}(\mu_F)$. Explicit calculation \cite{OMR} shows that the optimum choice is   $\mu_F=\mu_0=1.4M$. The stability of the prediction, using MSTW NLO PDFs \cite{MSTW}, is shown in Fig. \ref{fig:muF}(b).  For $Y \gapproxeq 3$, pure DGLAP PDF extrapolations become unreliable due to the absence of absorptive, ln$(1/x)$,..modifications. Rather, LHC data will provide a {\it direct} measure of PDFs in this low $x$ domain.

\section{Treatment of the infrared region in perturbative QCD}

Interestingly, a spin-off of the above study highlighted an inconsistency in the {\it conventional} treatment of the infrared region \cite{OMRir}. Again we use Drell-Yan as an example. For the main NLO subprocess we have
\begin{equation}
\frac{d \hat{\sigma}(gq\to q\gamma^* )}{d |t|}=\frac{\alpha^2 \alpha_s z}{9 M^2} \frac{1}{|t|} \left[( (1 - z)^2 + z^2) + z^2 \frac{t^2}{M^4} - 2 z^2 \frac{t}{M^2} \right],
\label{eq:nlo2}
\end{equation}
where $z=M^2/{\hat s}$ and $\sqrt{{\hat s}}$ is the incoming $gq$ c.m. energy. (Strictly speaking, $z$ is the ratio of the light-cone momentum fraction carried by the `daughter' quark to that carried by the `parent' gluon, $z=x^+_q/x^+_g$.)
In order to calculate the inclusive cross section $d {\sigma}/d M^2$, it seems that we have to integrate over $t$ starting from $t=0$. If this were necessary, then we would face an infrared divergency. 

However, we follow the procedure of the previous Section, which we call the {\it physical} approach. To avoid double-counting, we subtract the LO DGLAP-generated contribution. Then the remaining contribution of (\ref{eq:nlo2}) is \cite{OMRir}
\begin{equation}
\frac{d \hat{\sigma}^{\rm NLO}_{\rm rem}(gq\to q\gamma^* )}{d |t|}=\frac{\alpha^2 \alpha_s z}{9 M^2} \frac{1}{|t|} \left[[ (1 - z)^2 + z^2] ~\Theta(|t|-\mu_F^2) ~+~ z^2 \frac{t^2}{M^4} ~-~ 2 z^2 \frac{t}{M^2} \right].
\label{eq:nlo3}
\end{equation}
which has no singularity as $t \to 0$. The LO DGLAP evolution has accounted for all virtualities $|t|=k^2<\mu_F^2$; with the contribution of $k^2<Q^2_0$ hidden in the phenomenological input PDF at $Q^2=Q^2_0$.

On the other hand, the {\it conventional} prescription evaluates the inclusive cross section, $d\sigma/dM^2$, by integrating (\ref{eq:nlo2}) over the infrared divergency at $t=0$ using  $4+2\epsilon$ dimensional space to regularize the integral \cite{Alt2,ESW, Vrap}. Then the contribution from very small $t$ produces a $1/\epsilon$ pole, which is absorbed into the incoming PDF. 
%Simultaneously,
The conventional prescription is as follows. The same $gq \to q\gamma^*$ diagram, but now generated by LO DGLAP evolution, is considered; it gives an $1/\epsilon$ pole which cancels the corresponding 
 $1/\epsilon$ pole in hard matrix element (coefficient function).
However, we are left with  
 $\epsilon/\epsilon$ terms of infrared origin, which produce a 
 non-zero result as $\epsilon  \to 0$. Unlike the finite $\epsilon/\epsilon$ terms of ultraviolet origin, which can be treated as point-like counter-terms in the Lagrangian, the infrared $\epsilon/\epsilon$ contribution makes no physical sense in QCD theory, since the confinement eliminates any interaction at very large distances.
 
To be explicit, the {\it conventional} prescription gives 
\begin{equation}
\frac{M^2 d \hat{\sigma}^\mathrm{NLO}_\mathrm{rem} ( gq \rightarrow q \gamma^*)}{d M^2}
 = \frac{\alpha^2\alpha_s z}{9 M^2} 
\left\{ \left[ (1-z)^2 + z^2 \right] \ln \frac{(1-z)^2}{z} 
+ \frac{1}{2} + 3 z - \frac{7}{2} z^2 \right\},
\label{eq:con}
\end{equation}
which is to be compared with the result,
\begin{equation}
\frac{M^2 d \hat{\sigma}^\mathrm{NLO}_\mathrm{rem} ( gq \rightarrow q \gamma^*)}{d M^2}
 = \frac{\alpha^2\alpha_s z}{9 M^2} 
\left\{ \left[ (1-z)^2 + z^2 \right] \ln \frac{(1-z)}{z} 
+ \frac{1}{2} + z - \frac{3}{2} z^2 \right\},
\label{eq:phys}
\end{equation}
 obtained, from (\ref{eq:nlo3}), using the {\it physical} approach. We have checked that the spurious  $\epsilon/\epsilon$ contribution in the conventional approach is responsible for the difference,
\begin{equation}
\frac{\alpha^2\alpha_s z}{9 M^2} 
\left\{ \left[ (1-z)^2 + z^2 \right] \ln (1-z) 
+ 2 z (1 - z) \right\},
\end{equation}
between (\ref{eq:con}) and (\ref{eq:phys}).

 The message is that to get a reliable result we must use the {\it physical} approach and to subtract from (\ref{eq:nlo2}) the $1/t$ singularity exactly, so obtaining the non-singular expression (\ref{eq:nlo3}). Dimensional regularisation is not appropriate in the infrared region. As an example we have used the NLO coefficient function for Drell-Yan production.  However, the result applies more generally. It is relevant to other processes, such as deep inelastic scattering \cite{OMRir}. It is relevant to the NLO splitting functions.

It should be noted that the {\it physical} approach cannot be considered as an alternative factorization scheme. That is, the difference between the conventional and the physical coefficient functions cannot be compensated by a re-definition of the parton distributions (PDFs).
The corresponding (infrared) $\epsilon/\epsilon$
corrections in the splitting functions do not coincide with those that come from the re-definition of the PDFs.

\section*{Acknowledgements}
 This work was supported by the grant RFBR 11-02-00120-a
and by the Federal Program of the Russian State RSGSS-4801.2012.2;
and by FAPESP (Brazil) under contract 2012/05469-4.


\thebibliography{}

\bibitem{OMR} E.G. de Oliveira, A.D. Martin and M.G. Ryskin, Eur. Phys. J {\bf C72}, 2069 (2012).

\bibitem{MSTW} A.D. Martin, W.J. Stirling, R.S. Thorne and G. Watt, Eur. Phys. J. {\bf C63}, 189 (2009).

\bibitem{OMRir} E.G. de Oliveira, A.D. Martin and M.G. Ryskin, arXiv;1206.2223.
  
\bibitem{Alt2} G. Altarelli, R.K. Ellis and G. Martinelli, Nucl. Phys. {\bf B157}, 461 (1979).

\bibitem{ESW} See, for example, R.K. Ellis, W.J. Stirling and B.R. Webber, in {\it QCD and Collider Physics} (Cambridge Univ. Press, 1996) and refs. therein.

\bibitem{Vrap} C.~Anastasiou, L.J.~Dixon, K.~Melnikov and F.~Petriello,
  Phys.\ Rev.\  {\bf D69}, 094008 (2004).
\end{document}